\documentclass[final,10pt,twocolumn]{elsarticle}

 \usepackage{graphics}
\usepackage{graphicx}

\usepackage{amssymb}

\journal{Reuni\'on Nal. de F\'isica y Matem\'aticas 2011}

\begin{document}

\def\be{\begin{equation}}
\def\ee{\end{equation}}
\def\ba{\begin{eqnarray}}
\def\ea{\end{eqnarray}}

\begin{frontmatter}



\title{Classical equivalents for quantum eigenfunctions in energy representation for simple hamiltonian systems}
\author{H.~Hern\'andez-Salda\~na}
\address{Dpto. de Ciencias B\'asicas,
Universidad Aut\'onoma Metropolitana at Azcapotzalco,
Av. San Pablo 180, 02200, M\'{e}xico D.F., Mexico.}

\begin{abstract}
A calculation of the classical analogue for the quantum wave function and local density of states, in energy representation, is presented for simple Hamiltonian systems. Such analogous were proposed by M. V. Berry and A. Voros considering the intersection of energy shells of two systems as the only semiclassical object which can give support to eigenfunctions. One of them is the system under study and the other one is the “unperturbed” system used to express the wave functions, even in the case that both systems are not close. For simple systems and as for scalable ones analytical expressions are obtainable. In the present work we offer examples of both.
\end{abstract}

\end{frontmatter}

\section{Introduction}

Quantum classical correspondence has been an important issue since the creation of quantum theory. The general idea is expressed in Bohr's principle of correspondence. This subject is matter of current research in a wide class of aspects\cite{ar1}. Here we discuss the case of eigenfunctions. Several studies and approximation to the problem exists in the literature \cite{2} but we shall concentrate in the proposal by M. V. Berry \cite{3} and A. Voros \cite{4}. The argument is that the energy shell of the Hamiltonian systems is the only semi classical object who could give support to eigenfunctions. The practical development of such an idea was performed by Benet et al \cite{5}.

The specific form of such object is discussed in the Methodology section. This subject with some 
corrections \cite{6,7}, is expected to be the classical equivalent of quantum eigenfunctions, 
and we shall call them the Classical Eigenfunctions, CEF. In order to test this proposal and due 
to the nature of this object it is expected that a better quantum classical correspondence occurs 
in ergodic systems, naturally, the easiest way is to consider systems with a chaotically classical 
dynamics. Several systems have been considered in order to test the quantum classical correspondence, 
and, as a consequence the reliability of the CEF (see equation (\ref{eq:1}) below) as the classical 
equivalent for the quantum eigenfunctions. The systems considered are chaotic quartic oscillators \cite{5,6,7} 
and chaotic billiards \cite{8}, all them with good results. As pointed out in \cite{7}, a better 
fitting of CEF to quantum eigenfunctions is obtained when both hamiltonians, the system under study 
and that used as a basis, present a chaotic classical dynamics. This is a result that CEF depends on 
the ergodic properties of the systems. But what about integrable systems, how good could be the 
quantum classical correspondence in these systems?. Another important point is that in all previous 
works the CEFs were calculated numerically testing carefully the reliability, however no so much was 
clarified about the properties of them. An example of this is that in all the reported cases CEFs 
shown strong peaks, but nothing is discussed about their nature. 
Here, we start a small contribution on the subject.

\section{Methodology}

Looking for a practical formulation of Berry's and Voros' idea, L. Benet {\it etal} \cite{5} proposed that the quantity,
\be
\label{classical}\label{eq:1}
g({\cal E}, {\cal E}_0)=
A \int 
 \delta({\cal E}-H)
 \delta({\cal E}_0-H_0)
 {\rm d}{\bf p}{\rm d}{\bf q},
 \ee
could be the classical counterpart for the eigenfunctions of the hamiltonian $H$ 
at energy ${\cal E}$ in the basis of the hamiltonian $H_0$ with 
energies $ {\cal E}_0$. $A$ is a normalization constant which contains 
information about the quantum scale occupied by the state in phase space and 
$\delta$ is the Dirac's delta function. This quantity does not contain information 
about the quantum phase, neither is the solution of an eigenfunction equation. 
However, it corresponds to the intensity of a quantum eigenfunction $|\Psi \rangle$ in the $|\phi_0 \rangle$ basis; for  
\be
H |\Psi \rangle =  {\cal E}   |\Psi \rangle,
\ee
and 
\be
H_0 |\phi_0 \rangle =  {\cal E}_0   |\phi_0 \rangle,
\ee
respectively.
Evidence of this quantum classical correspondence for hamiltonian systems with a 
chaotic classical counterpart exists \cite{5,6,7,8}. The function $g$ represents the intersection of both energy shells, the $H$ 
and $H_0$ ones, and it is obtained in the same way as the average density of states though the Weyl formula 
(see, for instance, \cite{9}). It is important to recall that distributions like (\ref{eq:1}) or 
the Weyl approximation requires certain ergodicity in the system.

\begin{figure}[t!]
\includegraphics[width=\columnwidth]{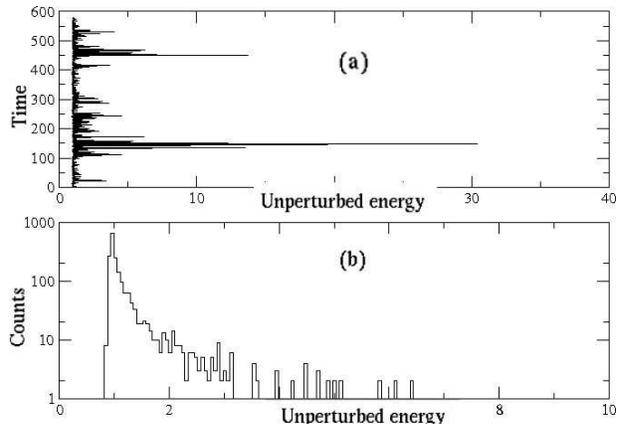}
\caption{(a) Time evolution of unperturbed energy at fixed system energy. (b) Histogram of unperturbed energy distribution for the case shown in (a). In log-linear scale. At the mixing time the distribution becomes (\ref{eq:1}). Units are arbitrary.} 
\label{Fig:RealTime}
\end{figure}

A way to understand the physical meaning of (\ref{eq:1}) go as follow: Consider a system $H$ composed by the sum of $H_0$ and 
a perturbation $\lambda W$, i.e., $H = H + \lambda W $. The perturbation is not necessarily small. At a fixed energy ${\cal E}$ of 
the system $H$, the energy is distributed between both parts, the unperturbed and the perturbation, 
hence the energy ${\cal E}_0$ is time dependent and its distribution defines (\ref{eq:1}). In other words, if we consider the 
classical solution of the test system and we evaluate these solutions in phase space in the nonperturbed hamiltonian, 
the corresponding energy is a time dependent variable. In Fig. 1a we show this process. As the time pass, 
a limit distribution appears as is shown in Fig. 1b. The distribution is the histogram of the projection into the 
unperturbed energy for a time sufficiently large. The time needed to reach such a distribution is called the mixing time. 
Meanwhile this picture is clear, a practical calculation of the distribution depends strongly on the initial conditions. 
Here and in \cite{5,6,7} the function in (\ref{eq:1}) is calculated using Montecarlo integration. This method allows an efficient calculation.

\subsection{ A small but important detail} 

Notwithstanding that (\ref{eq:1}) shows a good quantum classical correspondence in the cases presented in \cite{5,8}, 
expression (\ref{eq:1}) must be rectified in order to have a proper normalization. As we shall see, normalization depends on 
both energies ${\cal E}$, and ${\cal E}_0$ via their corresponding density of states. This detail becomes much more 
important as we consider a larger number of particles and the densities become larger as well. The definition proposed \cite{6,7} is
\ba
\begin{array}{lcl}\label{eq:2}
g_c({\cal E}, {\cal E}_0) &=&
\left( c/\rho({\cal E}) \rho_0({\cal E}_0) \right ) \\
 & \times & \int
 \delta({\cal E}-H)
  \delta({\cal E}_0-H_0)
   {\rm d}{\bf p}{\rm d}{\bf q},
\end{array}   
\ea
where $\rho({\cal E})$ and  $\rho_0({\cal E}_0)$ are the density of states for the 
perturbed and unperturbed systems respectively. They are calculated according to 
the Weyl approximation
\be
\rho(x) = (1/c) \int \delta (x-H) {\rm d}{\bf p}{\rm d}{\bf q},
\ee
for any of the hamiltonians. The constant $c$ is defined as $1/( 2\pi \hbar )^d$. Here, $d$ is 
the number of degrees of freedom of our systems. Even when calculations could be performed 
in three dimensional configuration space for any number of particles we shall 
concentrate in one dimensional systems and hence $d$ will be the number of particles 
considered. 

Now we shall explain how to arrive to correction (\ref{eq:2}) from (\ref{eq:1}). 
The arguments in order to arrive to expression (\ref{eq:2}) are as follow \cite{6}: 

Integrating (\ref{eq:1}) respect to ${\cal E}_0$ we obtain
\be
\int g({\cal E}, {\cal E}_0) {\rm d} {\cal E}_0 = \rho({\cal E}) ( 2\pi \hbar )^d
\ee
where we used the properties of the delta function. An integration on ${\cal E}$ gives 
as a result $\rho_0({\cal E}_0 ( 2\pi \hbar )^d) $.
The generalization to the continuum of the normalization rule of a quantum state on a 
discrete or a continuum spectrum requires an integration on the energy element $\rho_0({\cal E}_0) {\rm d} {\cal E}_0$. 
Hence, only (\ref{eq:2}) offers the correct normalized function that fulfills all these requirements. 

For practical porpoises we require a simplified version of (\ref{eq:2}). For simplicity we shall consider 
systems of one or several particles in one dimension, hence the parameters $d$ shall refer to the 
number of particles. In this section, as well, we shall refer to (\ref{eq:1}) instead of (\ref{eq:2}) in order 
to avoid to carry on the densities. In absence of magnetic filed, the integration of the momenta 
variables can be performed easily with the help of the property 
\be \label{eq:5}
\delta (x-f) \delta (x-g) =\delta(x-(f+g)/2) \delta(f-g),
\ee
for any scalar functions $f$ and $g$. This allows to rewrite the integrand of (\ref{eq:1}) as
\be \label{eq:6}
\delta(p^2/2-({\cal E}-V+{\cal E}_0-V_0)/2)
\delta({\cal E}-V-({\cal E}_0-V_0)),
\ee
for the potentials associated to each hamiltonian. Using spherical coordinates for the 
momenta in the argument of (\ref{eq:6}), the integral defining (\ref{eq:2}) can be rearranged as
\ba
\begin{array}{l}\label{eq:7}
g_c({\cal E}, {\cal E}_0) = B \int {\rm d}{\bf q} \delta({\cal E}-V-({\cal E}_0-V_0)) \\
  \times  \left ( {\cal E}-V+{\cal E}_0-V_0) \right ) ^{\frac{d-2}{2}} 
 \Theta(\left ( {\cal E}-V+{\cal E}_0-V_0) \right ).
\end{array}
\ea

Here $\Theta(\cdot)$ is the step function and appears in order to keep the positivity in the square root argument. $B$ stands for
\be
B= 2 \pi^{d/2}/\left( (2\pi \hbar)^d \Gamma(d/2) \rho({\cal E}) \rho_0({\cal E}_0). 
\right )
\ee
Where $\Gamma$ stands for the Gamma function. This constant contains the result from the angular part in the momentum integral. 
We shall use extensively (\ref{eq:7}) for the calculations presented in this work. 

Property (\ref{eq:5}) can be obtained straightforwardly using the sequence of gaussians with limit a 
Dirac delta function. Assuming that the sequence if $n$ gaussians is taken at the same time for each case, 
we can arrange properly the argument in both gaussians and take the limit to infinity in order to obtain 
the deltas again(see appendix in \cite{6}).

\subsection{The systems}

 In order to understand the use and the properties of (\ref{eq:2}) we shall consider several simple systems and one not so 
simple in order to contrast our results. The selected systems are a general one dimensional Hamiltonian in their 
own basis, the free particle in its own basis, a quartic oscillator in the harmonic basis and, at the end, a chaotic 
several particles Hamiltonian in the basis of the quartic single particle basis.

\section{Results}

\subsection{Simple special cases in one dimension}

 Here we apply the considerations discussed in previous section. We discuss first some very simple examples in one dimension 
for one particle. In subsection B we shall discuss the cases with more than one particle. 

\subsubsection{ Equal Hamiltonians in $d$ dimensions}

Eigenfunctions of quantum systems written in their own basis must be Kronecker's or Dirac's deltas. 
This is the case for the classical counterpart as well. To show this we use (\ref{eq:1}) and we consider as an 
independent variable the whole Hamiltonian $H$, hence we have
\ba
\begin{array}{lcl}
g({\cal E}, {\cal E}_0) & = &   \int {\rm d}{\bf q} {\rm d}{\bf q} \delta( {\cal E}-H
\delta({\cal E}_0-H_0)), \\
 & = & \delta( {\cal E}- {\cal E}_0) (2\pi \hbar) ^{\frac{d}{2}} \rho({\cal E}).
\end{array}
\ea
We used property (\ref{eq:5}) in order to get the last line. The CEF could be written as 
\be \label{eq:10}
g_c({\cal E}, {\cal E}_0) = \delta( {\cal E}- {\cal E}_0)  \rho_0({\cal E}_0).
\ee
Notice that, as expected, these results do not depend on the potentials used or the system's dimensionality.

\subsubsection{ Free particle in its own basis}

 This a particularly simple case, the free particle in one dimension with $H=H_0=p^2/2$. 
Using the previous dimension with result (\ref{eq:10}) we only require to evaluate the density of states $\rho_0({\cal E}_0)$, giving
\ba
\begin{array}{lcl}
\rho_0({\cal E}_0) &=& (1/2\pi \hbar) \int \delta(p^2/2-{\cal E}_0){\rm d}p {\rm d}q \\
 &=& (1/2\pi \hbar) (L/\sqrt{2 {\cal E}}).
\end{array}
\ea
Where we performed the required integral in $q$ on a box of size $L$ and we take the limit to infinity at the end. 
The final result is 
\be
g_c({\cal E}, {\cal E}_0) = (2\pi\hbar \sqrt{2 {\cal E}} ) \delta ({\cal E} -{\cal E}_0).
\ee
The dependence on $L$ disappears when the normalizations is performed. The same result is obtained by direct integration.

\subsubsection{ Quartic oscillator in the harmonic basis.} \label{qoinhb}

 Now we consider the perturbed Hamiltonian in the $H = p^2/2 + \beta q^2 + \alpha q^4$  basis of the 
harmonic hamiltonian, i.e.,  $H = p^2/2 + \beta q^2 $. Using (\ref{eq:7}) with $d=1$ to  this case we have 
\be \label{eq:13}
g_c \propto \int {\rm d}q \delta ( f(q) ) /
\sqrt{{\cal E} - (\beta q^2 + \alpha q^4)+({\cal E}_0-\beta q^2)},
\ee
with
\be
f(q)= {\cal E} - (\beta q^2 + \alpha q^4)-({\cal E}_0-\beta q^2),
\ee
 and we dropped the unit step function for sake of clarity. Eq. (\ref{eq:13}) can be re-written as
\be  \label{eq:15}
g_c \propto \int {\rm d}q \delta({\cal E} -{\cal E}_0 -\alpha q^4 )/
\sqrt{{\cal E}+{\cal E}_0 -(2\beta q^2 + \alpha q^4)}.
\ee
We require the roots in the argument of the delta function, that is
\be \label{eq:16}
q = \sqrt[4]{({\cal E} - {\cal E}_0)/\alpha}.
\ee
Using the property of delta functions
\be
\delta(f(x)) = \sum_i \delta (x-x_i) /|f'(x_i)|,
\ee
on (\ref{eq:15}), we obtain
\be  \label{eq:18}
g_c\propto \left(({\cal E} - {\cal E}_0)/\alpha \right)^{-3/4} 
 \left({\cal E}_0 - \beta \sqrt{({\cal E} - {\cal E}_0)/\alpha}\right)^{-1}.
\ee

The density of states is constant for both Hamiltonians, the unperturbed one is proportional to 
the mean level density, which, in this case, is constant. 
(The spectrum of the one dimensional harmonic oscillator is equidistant). 
The density of states for the perturbed case can be calculated explicitly with the help of 
symbolic calculus software, but at the end it have a is a number. From here it is clear 
that $g_c\left(({\cal E} ,{\cal E}_0\right)$ have a  divergence when both energies are equal. 
Notice, as well, that another classically forbidden region appears when the argument the 
asymmetry in the root argument 
${\cal E}_0 - \beta \sqrt{({\cal E} - {\cal E}_0)/\alpha}$ becomes negative, that is, 
it corresponds to an asymmetrical function. In Fig. 2 we show a graph of this case. 
In broken red line appears the classical analogue (\ref{eq:18}) for the energy corresponding to 
the state number 756 of the perturbed system. In black points the corresponding quantum result. 
An average on windows of size 30 was performed in order to smooth the quantum oscillations, 
the result is shown in blue crosses. The agreement is notable. 
A shift (not shown) of the classical eigenfunction is near of the quantum case envelope.

Notice that the non-oscillating tails in the quantum case, shown in Fig 2, correspond to the 
tunnel effect to the classically forbidden zone. The quantum case was calculated using the 
diagonalization of the Hamiltonian matrix in the unperturbed basis using 4000 states.

\begin{figure}[t!]
\includegraphics[width=0.9\columnwidth,angle=0]{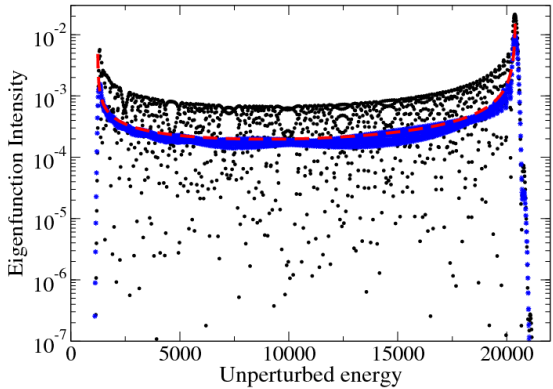}
\caption{(Color on-line) 
Eigenfunction intensities in semi log scale for the state 756 of the single particle quartic oscillator in the harmonic basis ( black pints), classical equivalent for the same case (broken red line). We add an averaged version of the quantum eigenfunction.
}
\end{figure}

\subsection{Not so simple cases} \label{notsosimple}

Analogues like (\ref{eq:7}) are builded up for complex systems where a certain amount of 
ergodicity and perhaps mixing exists. For such cases quantum classical correspondence has 
been good \cite{5,6,7,8}, here we shown another example. The Hamiltonian considered is equal to those appearing in \cite{6,7}
\be \label{eq:19}
H =T 
+ \sum \alpha{ q}_i^4
+\beta\sum{q}_i^2 { q}_j\!^2
+ \gamma\sum({ q }_i^3{q}_j+
 { q }_i{ q }_j^3),
 \ee
with the following values for the  parameters  $\alpha =10.0$, $\beta=-2.1$ and $\gamma= -2.2$, and the number of particles is $d=4$. 
Here, $T$ stands for the kinetic energy. The numerical calculation was performed diagonalizing a $10000 \times 10000$ 
matrix with the described method in \cite{10} and a proper basis cutoff. The chaoticity was 
tested calculating the maximum Lyapunov exponent for a large variety of initial conditions. 
In Fig. 3 we plot an average of 100 eigenfunctions around the state number 500. 
The quantum classical correspondence is good but a shoulder of quantum nature appears. 
This shoulder causes that the fit at the tail fails as a result of the normalization rule. 
Notice that the classical analogue present a strong peak at the “resonance” energy meanwhile the quantum 
counterpart present a large intensity at the surrounding but not precisely at that energy. 
In fact, the average of quantum
eigenfunctions was performed shifting the eigenfunctions in order that the maximum intensity 
coincide in each case. This is possible due that we are considering eigenstates sufficiently high in the energy spectrum.

\begin{figure}[t!]
\includegraphics[width=0.9\columnwidth,angle=0]{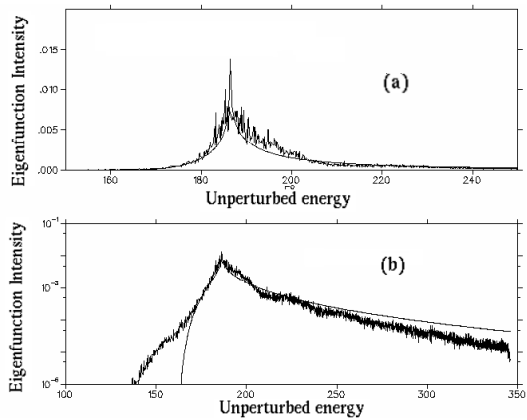}
\caption{ (a) Average intensities for the state number 500 over 101 eigenfunctions around it (fluctuating curve) and their corresponding classical analogous (smooth curve). (b) Same as before but in semi log scale.
}
\label{Fig:Chambers2006}
\end{figure}

\section{Discussion}

Even when the reliability of the numerical calculations is achieved, several characteristics of CEF (\ref{eq:7}) remains obscure, 
that is the motivation of this simple calculations. As explained in previous sections, the analytical results depends 
in depth on the nature of the potentials used. However some general features could be discussed. First, the square root 
and the positivity of the arguments is of general nature for conservative systems, hence such characteristics will 
remain even when we use several particle Hamiltonians in three dimensions. These characteristics could be important 
since eigenfunctions equivalents could give a faster estimation for some wave functions of interest in atomic or 
molecular physics. Additionally, the zeros of the arguments in (\ref{eq:2}) or (\ref{eq:7}) define the border to the 
classically forbidden zone and the turning points. Second, one of main problems in order to obtain completely the 
integrals involved in (\ref{eq:7}) is that we require the knowledge of the zeros of a complicated multidimensional 
function, and express them in a generalization of (\ref{eq:16}).

 Another interesting point to discuss is about the CEF behavior ${\cal E}_0$ near ${\cal E}$. 
From the numerical calculations this behavior at point seems like a divergence and not a smooth peak. 
The examples presented here have a divergence at this point. Hence for all these cases a divergence peakshould appears in the case that the unperturbed and the perturbed energies coincide. 
Others divergences associated with 
classical turning points must appear in the evaluation of (\ref{eq:7}) as appears in 
the case presented in \ref{qoinhb}. 
How CEF looks like for a larger number of particles in this case is not obvious since 
degenerancies appear in 
the several particles spectrum, which invites to look for another example where the generalization to several particles and simplicity of analytical calculations can be applied. 

For a large number of particles, the limit expression of (\ref{eq:7}) tend to be a gaussian like 
form [6]. This shape is the expected one for general eigenfunctions like those generated by 
gaussian ensembles of matrices \cite{11}. How eigenfunctions change this behavior is not clear, 
but even for the numerical calculations presented here and those in progress for a larger number 
of particles the peaks appear as a divergence.

Up to now we deal with eigenfunctions, that is, we consider CEF (\ref{eq:2}) and (\ref{eq:7}) as a function of unperturbed energy at fixed ${\cal E}$ . For the local density of states the oposit is considered, (\ref{eq:2}) 
and (\ref{eq:7}) become functions of ${\cal E}$ instead of ${\cal E}_0$. This case is of relevance for experimental comparison. Due to the symmetry in (\ref{eq:2}) we can use, {\it mutatis mutandis}, the same expressions as before, including (\ref{eq:7}). In Fig. 4 we shown the LDOS calculation for the same case presented in section \ref{notsosimple}. In smooth line the classical counterpart obtained from (\ref{eq:2}) and in strongly oscillating line the corresponding quantum case averaged in a similar way as before. Notice that the quantum classical correspondence is much better fitted, and the shoulder is not present. 

\begin{figure}[t!]
\includegraphics[width=1.0\columnwidth]{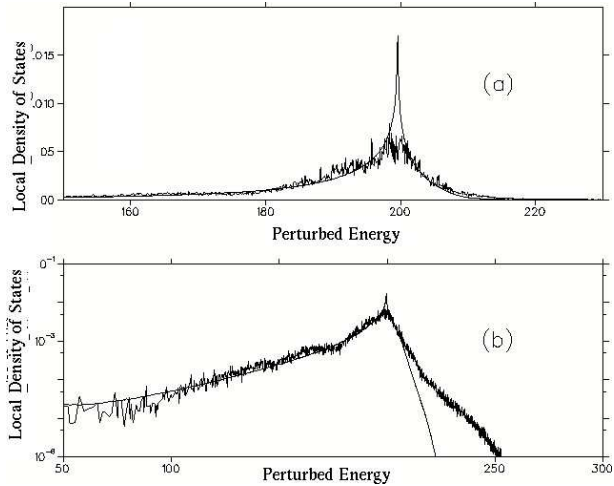}
\caption{Local Density of States for the anharmonic oscillator presented in the text. Classical analogue in smooth curve.}
\end{figure}

\section{Conclusions}

Classical analogues for eigenfunctions, Local Density of States and occupation number 
statistics have been proposed in \cite{5,6,7,8} with a good quantum classical 
correspondence for few particle anharmonic hamiltonian and for a chaotic billiard. 
Evaluation of such analogues has been performed numerically and some characteristics 
remain obscure. Here we started a program in order to understand general properties of 
those quantities, like divergences, ergodicity and tunnel effect. For that matter we 
evaluate analytically the expression for the classical equivalents for very simple cases. 
The cases were selected for simplicity and in order to identify the causes of divergence 
without the use of a numerical calculation. The cases presented here shown that the 
numerical eigenfunctions calculated have a divergence when the energy coincides with 
the perturbed one. An interesting point is the limit where a large number of particles 
is considered. In such a case eigenfunctions shown a rounded peak at the “resonance” 
energy and a gaussian form, unfortunately the harmonic oscillators considered in the 
present work have a lot of degenerancies and them avoid to understand the limit, even 
when in the case presented here have a notable good quantum classical correspondence for 
one single particle and without the requirement of ergodicity. 

As a final remark we should stress that classical eigenfunction analogue (\ref{eq:2}) is well situated for a fast calculation of average eigenfunctions in quantum complex systems.

\section*{Aknowlegments}

This work received financial support from PROMEP/SEP, project 2115/35621.

\end{document}